\documentclass{amsart}
\usepackage{eurosym}
\usepackage{amsfonts}
\usepackage{amsmath}
\usepackage{amssymb}
\usepackage{graphicx}
\usepackage{amscd}
\usepackage[utf8]{inputenc}

\newtheorem{theorem}{Theorem}
\theoremstyle{plain}

\newtheorem{lemma}{Lemma}

\numberwithin{equation}{section}

\begin{document}
\title[Asymptotic theory of Takayama's index]{Asymptotic Theory and Statistical Decomposability gap Estimation for Takayama's Index}
\author{$^{(1)}$ Mergane P.D.}
\author{$^{(1)}$ Haidara C.M.}
\author{$^{(2)}$Cheikh Seck}
\author{$^{(1,3,4)}$Lo G.S.$^{\ast}$}

\begin{abstract} In the spirit of recent asymptotic works on the General Poverty Index (GPI) in the field of Welfare Analysis, the asymptotic representation of the non-decomposable Takayama's index, which has failed to be incorporated in the unified GPI approach, is addressed and established here. This representation allows also to extend to it, recent results of statistical decomposability gaps estimations. The theoretical results are applied to real databases. The conclusions of the undertaken applications recommend to use Takayama's index as a practically decomposable one, in virtue of the low decomposability gaps with respect to the large values of the index.  

\bigskip
\bigskip \noindent $^{(1)}$ \textit{Pape Djiby Mergane (merganedjiby@gmail.com), Cheikh Mohamed Haidara (chheikhh@yahoo.com), Gane Samb Lo (gane-samb.lo@ugb.edu.sn)}. LERSTAD, Gaston Berger University, Saint-Louis, Senegal.\newline

\noindent $^{(2)}$ \textit{Cheikh Tidiane Seck (cheikh.seck@uadb.edu.sn)}. University Alioune Diop of Bambey, SENEGAL.\newline

\noindent $^{(3)}$ LSTA, Pierre and Marie Curie University, Paris VI, France.\newline
\noindent $^{(4)}$ \textit{Gane Samb Lo (gslo@aust.edu.ng)}. AUST - African University of Sciences and Technology,
Abuja, Nigeria\newline

\noindent \textit{$^{\ast}$ Corresponding author}. Gane Samb Lo. Email :
gane-samb.lo@edu.ugb.sn, ganesamblo@ganesamblo.net, gslo@aust.edu.ng.\\
\textit{Permanent address} : 1178 Evanston Dr NW T3P 0J9,
Calgary, Alberta, Canada.
\end{abstract}

\subjclass[2000]{Primary 62G30; 62E20; 62H12; 62G15. Secondary 91B15; 91B02; 91B82}
\keywords{Welfare index; Asymptotic Representation; Asymptotic Laws; Statistical estimation of decomposability; Welfare Axiomatic; Functional Gaussian Process; Gaussian Fields}

\maketitle

.

\section{Introduction} \label{sec1}

\subsection{General introduction, motivations and objectives} \label{subsec11}
In this paper, we are concerned with the asymptotic theory of the Takayama (\cite{takayama}) welfare statistic and the estimation of its decomposability gap. Let us begin to define it. Let $X_1,$ $X_2$, etc., be independent observations of a non-negative random variable $X$ with cumulative distribution function (cdf) $F_{(1)}$, all of them defined on the same probability space $(\Omega, \mathcal{A}, \mathbb{P})$, and for each $n\geq 1$, let $\mu_n$ be the sample mean.\\

\noindent In Welfare Analysis, a poverty line is set as the lowest income under which an individual is declared poor. The number of poor individuals in the sample is denoted by $Q_n$. Now, we consider for each $n \geq 1$, the Takayama statistic as defined by

\begin{equation}
T_n(X) = 1+\frac{1}{n}-\frac{2}{\mu _{n}n^{2}}\sum_{j=1}^{q}\left(n-j+1\right) X_{j,n}. \label{takayamaEmp}
\end{equation}

\noindent We will see in Theorem \ref{theo1} that, under suitable conditions, Takayama's index (\ref{takayamaEmp}) converges in probability to  

\begin{equation}
T = 1 - \frac{2}{\mu} \int y(1-F_{(1)}(x))\mathbf{1} \!\!\mathrm{I}_{(x\leq Z)} dF_{(1)}(x). \label{takayamaExact}
\end{equation}

\bigskip \noindent as $n\rightarrow +\infty$. Accordingly, the number $T$ is defined as the Takayama parameter of the \textit{cdf} $F_{(1)}$.\\

\noindent This statistic has been extensively studied by many authors from the axiomatic point of view. Indeed, Welfare Analysis researchers investigate the quality of an index with respect to a number of desired axioms. In that sense, the review paper of \cite{zheng} is a useful reference. We are going to quote \cite{zheng}, just to highlight its importance and, for this reason,  we will not enter into the details of the meanings of these axioms. According to \cite{zheng}, Takayama's measure satisfies the following axioms : Focus, symmetry, replication invariance, continuity,
minimal transfer, restricted continuity, nonpoverty growth, normalization. And it fails to fulfill the others : weak transfer, progressive transfer,
decomposability, regressive transfer, weak transfer sensitivity, subgroup consistency, weak monotonicity, strong monotonicity.\\

\noindent With respect to its relation with the Gini inequality, the Takayama measure is a smoother translation of the Gini coefficient; but such advantage is obtained at a substantial cost. (See \cite{zheng}). As Takayama himself admitted, his measure may violate the monotonicity axioms, which is a serious drawback. It violates every axiom that the Sen measure fails to satisfy except continuity and
replication invariance. The claim by Takayama that his estimator is superior to that of Sen has been challenged..".\\

\noindent One of the most desired axiom of a welfare measure is the decomposability one.  Let us explain this property at a statistical level.\\

\noindent Suppose that we are monitoring some index $I$ over a given population of size $N$. When $I$ is applied to the whole population, we may use the notation $I=I_{N}$. In a large population subjected to a number of  inequalities between areas and in which there are groups with specific features at the exclusion of the others, public policy efficiency usually requires to target disadvantaged areas or groups and to implement therein strong strategies aimed at improving the status of this group in relation to a given pattern (for example poverty, health covering, education level, etc.), monitored by the index $I$. In such a case, the population is divided into sensitive $K$ subgroups of interest $S_{1},...,S_{K}$ of respective sizes $N_i$, $i\in \{1,...,K\}$, and the studied behavior is followed up by an index, say $I$, taking the values $I_{N_{i}}(i)$ in each subgroup $S_i$, $i\in \{1,...,K\}$.\\

\noindent The index $I$ is said to be decomposable if we may express the \textit{global} index on the whole population with respect to the partial indices at the subgroup level as follows, that is

\begin{equation}
I_N= \sum_{1\leq i \leq K} \frac{N_i}{N} I_{N_i}(i). \label{formule.decom}
\end{equation}

\noindent Formula (\ref{formule.decom}) offers the practical and comfortable latitude to work at the local level with the possibility to recompose the global index at the global level. This explains why decomposable indices are so preferred, in particular the Foster-Greer-Thorbecke (\cite{fgt}) index of index $\alpha \geq 0$

$$
FGT_n(\alpha)=\frac{1}{n} \sum_{1\leq i \leq n} \max\left(\frac{Z-X_i}{Z}, \ 0\right)^{\alpha}, \ \ \alpha\geq 0.
$$

\noindent The problem is that some the most interesting measures are not decomposable, in particular the weighted ones. Indeed, successful policies require to target disadvantaged or vulnerable groups. For example, suppose that we are dealing with poverty. A measure that counts all poor individuals with the same weight is less interesting than another that puts bigger weights to poorer individuals. A variation of such an index in the good direction tends to be negligible if the less poor individual behave better, and to be noticeable if the poorer individuals among the poor become better off.\\

\noindent Our problematic is to keep using weighted measures like Sen \cite{sen}, Kakwani \cite{kakwani}, Shorrocks \cite{shorrocks} and Takayama measures, to cite a few, and yet, to have a quick approach to report the global situation. The solution resides certainly in the estimation of the decomposability gap.

\begin{equation}
g_N=I_N- \sum_{1\leq i \leq K} \frac{N_i}{N} I_{N_i}. \label{formule.decom}
\end{equation}

\bigskip
\noindent We will see later that we will be able to estimate this gap. Then we will be able to work at a local level and to report the global index in accurate confidence interval.\\

\noindent Recently, Haidara and Lo \cite{haidara-lo} motivated the estimation of decomposability gap of non-decomposable measures in the sense described above. But the results in \cite{haidara-lo}, although including almost all the known measures, ignored the Takayama for the main reason that this latter is not based on the poverty deficits $Z-X_{j,n}$ nor on the relative poverty deficits $(Z -X_{j,n})/Z$, $1\leq j \leq Q_n$, where $Q_n$ is the number of poor individuals, numbered from $1$ to $j$, and $Z$ is the poverty line.\\

\noindent This is the main motivation of studying the Takayama's index with respect to two directions :\\

\noindent (a) Provide a full asymptotic theory of Takayama's index, in parallel with that of General Poverty Index (GPI) in which its fails to a part of it. This asymptotic theory is based on the use of the functional empirical process and provides the results in the form of asymptotic representations.\\

\noindent (b) Based on the results of Point (a), the same functional empirical process is used again to handle the decomposability gap.\\

\noindent Our best achievements are the complete description of the asymptotic distribution of the generalized Takayama measure using the functional empirical process and an auxiliary empirical process we name \textit{residual one} in \cite{sall-lo}, and the statistical estimation of decomposability gap.\\

\noindent The rest of paper is organized as follows. In the rest of this Section \ref{sec1}, we describe the probability space on which the proofs will take place. In Section \ref{sec2}, an asymptotic representation theorem for the generalized Takayama index is stated and proved. In Section \ref{sec3}, the statistical estimation of the decomposability gap will be fully developed. 
In Section \ref{sec3} data-driven applications using real data are provided. A conclusion section will end the paper.\\

\subsection{Notations and Probability Space} \label{subsec12} $ $\\

\noindent We are going to describe the general Gaussian field in which we present our results. Indeed, we use a unified approach when dealing with the asymptotic theories of the welfare statistics. It is based on the Functional Empirical Process (\textit{fep}) and its Functional Brownian Bridge (\textit{fbb}) limit. It is laid out as follows.\\

\noindent When we deal with the asymptotic properties of one statistic or index at a fixed time, we suppose that we have a non-negative random variable of interest which may be the income or the expense $X$ whose probability law on $(\mathbb{R},\mathcal{B}(\mathbb{R}))$, the Borel measurable space on $\mathbb{R}$, is denoted by $\mathbb{P}_{X}.$ We consider the space $\mathcal{F}_{(1)}$ of measurable real-valued functions $f$ defined on $\mathbb{R}$\ such that \begin{equation*}
V_{X}(f)=\int (f-\mathbb{E}_{X}(f))^{2}d\mathbb{P}_{X}=\mathbb{E}(f(X)-\mathbb{E(}f(X)))^{2}<+\infty ,
\end{equation*}

\noindent where 
\begin{equation*}
\mathbb{E}_{X}(f)=\mathbb{E}f(X).
\end{equation*}

\bigskip \noindent On this functional space $\mathcal{F}_{(1)},$\ which is endowed with the $L_{2}
$-norm
\begin{equation*}
\left\Vert f\right\Vert _{2}=\left( \int f^{2}d\mathbb{P}_{X}\right) ^{1/2},
\end{equation*}

\noindent we defined the Gaussian process $\{\mathbb{G}_{(1)}(f),f\in \mathcal{F}_{(1)}\},$
which is characterized by its variance-covariance function

\begin{equation*}
\Gamma_{(1)}(f,g)=\int^{2}(f-\mathbb{E}_{X}(f))(g-\mathbb{E}_{X}(g))d\mathbb{P%
}_{X},(f,g)\in \mathcal{F}_{(1)}^{2}.
\end{equation*}
 
\noindent This Gaussian process is the asymptotic weak limit of the sequence of functional empirical processes (fep) defined as follows. Let $X_{1},X_{2},...$ be a sequence of independent copies of $X$. For each $n\geq 1,$ we define by 
\begin{equation*}
\mathbb{G}_{n,(1)}(f)=\frac{1}{\sqrt{n}}\sum_{i=1}^{n}(f(X_{i})-\mathbb{E}%
f(X_{i})),f\in \mathcal{F}_{(1)},
\end{equation*}

\noindent as the functional empirical process associated with $X$. \noindent Denote by $\ell^{\infty }(T)$ the space of real-valued bounded functions defined on $T=\mathbb{R}$ equipped with its uniform topology. In the terminology of the weak convergence theory, the sequence of objects $\mathbb{G}_{n,(1)}$  weakly converges to $\mathbb{G}_{(1)}$ in $\ell^{\infty}(\mathbb{R})$, as stochastic processes indexed by  $\mathcal{F}_{(1)}$, whenever it is a Donsker class. The details of this highly elaborated theory may be found in Billingsley \cite{billingsley}, Pollard \cite{pollard}, van der Vaart and Wellner \cite{vaart} and similar sources.\\

\noindent But, for our purposes, we only need the convergence in finite distributions which is a simple consequence of the multivariate central limit theorem, as described in Chapter 3 in Lo \cite{wcia-srv-ang}.\\

\noindent We also have to use the Renyi's representation of the random variable $X$ of interest by means of the  cumulative distribution function (\textit{cdf}) $F_{(1)}$ as follows
$$
X=_{d}F_{(1)}^{-1}(U),
$$

\noindent where $U$ is a uniform random variable on $(0,1)$, $=_{d}$ stands for the equality in distribution and $F^{-1}$ is the generalized inverse of $F$, defined by

$$
F_{(1)}^{-1}(s)=\inf \{x, F_{(1)}(x)\geq s\}, \ s \in (0,1).
$$

\noindent Based on these representations, we may and do assume that we are on a probability space $(\Omega,\mathcal{A},\mathbb{P})$ holding a sequence of independent $(0,1)$-uniform random variables $U_1$, $U_2$, ..., and the sequence of independent observations of $X$ are given by 

\begin{equation}
X_{1}=F_{(1)}^{-1}(U_1), \ \ X_{2}=F_{(1)}^{-1}(U_2), \ \ etc. \label{repRenyi}
\end{equation}

\noindent
\noindent For each $n\geq 1$, the order statistics of $U_1,...,U_n$ and of $X_1,...,X_n$ are denoted respectively by $U_{1,n}\leq \cdots \leq U_{n,n}$ and $X_{1,n}\leq \cdots \leq X_{n,n}$.\\

\noindent To the sequences of $(U_n)_{n\geq 1}$, we also associate the sequence of real empirical functions

\begin{equation}
\mathbb{U}_{n,(1)}(s)=\frac{1}{n} \#\{i,1\leq i \leq n, \ U_i \leq s\}, \ s\in(0,1) \ n\geq 1 \label{empiricalfunctionU}
\end{equation}

\bigskip \noindent and the sequence of real uniform quantile functions
\begin{equation}
\mathbb{V}_{n,(1)}(s)=U_{1,n}1_{(0\leq s \leq 1/n)}+\sum_{j=1}^{n}U_{j,n}1_{((i-1)/n\leq s \leq (i/n))}, \ s\in(0,1), \ n\geq 1 \label{quantilefunctionU}
\end{equation}

\noindent and next, the sequence of real uniform empirical processes 
\begin{equation}
\alpha_{n,(1)}(s)=\sqrt{n}(\mathbb{U}_{n,(1)}-s), \ s\in(0,1) \ n\geq 1 \label{empiricalprocessU}
\end{equation}

\bigskip 
\noindent and the sequence of real uniform quantile  processes

\begin{equation}
\gamma_{n,(1)}(s)=\sqrt{n}(s-\mathbb{V}_{n,(1)}), \ s\in(0,1) \ n\geq 1. \label{quantileprocessU}
\end{equation}

\bigskip \noindent The same can be done for the sequence $(X_n)_{n\geq 1}$, and we obtain the associated sequence of real empirical processes  

\begin{equation}
\mathbb{G}_{n,r,(1)}(x)=\sqrt{n} \left( \mathbb{F}_{n,(1)}(x)-F_{(1)}(x)\right), \ x\in \mathbb{R}, \ n\geq 1, \label{empiricalprocess}
\end{equation}

\bigskip \noindent where 

\begin{equation}
\mathbb{F}_{n,(1)}(s)=\frac{1}{n} \#\{i,1\leq i \leq n, \ X_i \leq s\}, \ x\in \mathbb{R} \ n\geq 1, \label{empiricalfunction}
\end{equation}

\noindent is the associated sequence of empirical functions,  and the associated sequence of quantile processes 

\begin{equation}
\mathbb{Q}_{n,(1)}(x)=\sqrt{n} \left( \mathbb{F}^{-1}_{(n),(1)}(s) - F^{-1}(s) \right), \ s\in (0,1), \ n\geq 1 \label{quantileprocesses}
\end{equation}

\bigskip
\noindent where 

\begin{equation}
\mathbb{F}^{-1}_{(n),(1)}(s)=X_{1,n}1_{(0\leq s \leq 1/n)}+\sum_{j=1}^{n}X_{j,n}1_{((i-1)/n\leq s \leq (i/n))}, \ s\in(0,1), \ n\geq 1, \label{quantilefunction}
\end{equation}

\noindent is the associated sequence of quantile processes.\\

\noindent By passing, we recall that $\mathbb{F}^{-1}_{(n),(1)}$ is actually the generalized inverse of $\mathbb{F}_{(n),(1)}$.  In virtue of the representation (\ref{repRenyi}), we have the following remarkable relations :

\begin{equation}
\mathbb{G}_{n,r,(1)}(x)=\alpha_{n,(1)}(F_{(1)}(x)), \ x\in \mathbb{R} \label{EmpEmpprocess}
\end{equation}

\bigskip \noindent and

\begin{equation}
\mathbb{Q}_{n,(1)}(x)=\sqrt{n}\left( F^{-1}_{(1)}(\mathbb{V}_{n,(1)}(s))- F^{-1}_{(1)}(s)\right) \ s\in (0,1), \ n\geq 1, \label{QQprocess}
\end{equation}

\bigskip \noindent We also have the following relations between the empirical functions and quantile functions

\begin{equation}
\mathbb{F}_{n,(1)}(x)=\mathbb{U}_{n,(1)}(F_{(1)}(x)), \ x\in \mathbb{R} \label{EEFunction}
\end{equation}

\bigskip \noindent and

\begin{equation}
\mathbb{F}^{-1}_{n,(1)}(s)=F^{-1}_{(1)}(\mathbb{V}_{(n),(1)}(s)), \ s\in (0,1), \ n\geq 1. \label{QQFunction}
\end{equation}

\bigskip
\noindent As well, the real and functional empirical processes are related as follows

\begin{equation}
\mathbb{G}_{n,r,(1)}(x)=\mathbb{G}_{n,(1)}(f_{x}^{\ast}), \ \alpha_{n,(1)}(s)=\mathbb{G}_{n,(1)}(f_s),  \ s \in (0,1)  \ x \in \mathbb{R}, \ n\geq 1, \label{empiricalprocessRealFonct}
\end{equation}

\bigskip \noindent where for any $x \in \mathbb{R}$, $f_{x}^{\ast}=1_{]-\infty,x]}$ is the indicator function of $]-\infty,x]$ and for $s \in (0,1)$, $f_s=1_{[0,1]}$.\\
 
\bigskip 
\noindent To finish the description, a result of Kiefer-Bahadur (See \cite{bahadur66}) that says that the addition of the sequences of uniform empirical processes and uniform quantile processes (\ref{empiricalprocessU}) and (\ref{quantileprocessU}) is asymptotically, and uniformly on $[0,1]$, zero in probability, that is

\begin{equation}
\sup_{s\in [0,1]} \left\vert \alpha_{n,(1)}(s)+\gamma_{n,(1)}(s) \right\vert =o_{\mathbb{P}}(1) \text{ as } n\rightarrow +\infty. \label{bahadurRep}
\end{equation}

\noindent This result is a powerful tool to handle the rank statistics when our studied statistics are $L$-statistics.\\

\noindent All the needed notation are now complete and will allow the expression of the asymptotic theory we undertake here.\\

\section{The asymptotic behavior of Takayama's statistic} \label{sec2}

Let us introduce the following notation. The mean value of $X$ is finite and is denoted by

$$
\mu =\mathbb{E}(X)
$$

\bigskip \noindent For a measurable numerical function $f$, we set 
\begin{equation*}
\mathbb{P}_{X}(f)=\int f(x)dF_{(1)}(x)
\end{equation*}

\noindent and
\begin{equation*}
\mathbb{P}_{n}(f)=n^{-1}\sum_{j=1}^{n}f(X_{j}).
\end{equation*}

\noindent Let us define 
\begin{equation*}
\mu _{n}=\mathbb{P}_{n}(I_d),
\end{equation*}

\bigskip \noindent where $I_d$ is the identity application on $\mathbb{R}$. Fix also, for $y\in 
\mathbb{R}_+{\setminus\left\{0\right\}},$%
\begin{equation*}
\ell (x)=x\,\mathbf{1} \!\!\mathrm{I}_{(x\leq Z)},
\end{equation*}

\begin{equation*}
h(x)=x(1-G(x))\mathbf{1} \!\!\mathrm{I}_{(x\leq Z)},
\end{equation*}

\begin{equation*}
g(x) = 2\,\left(\mathbb{P}_{X}(h)\, \mathbb{E}^{-2}\left(X\right)\,x - 
\mathbb{E}(X)^{-1}\,h(x) \right),
\end{equation*}

\begin{equation*}
q(x)=-2\mathbb{E}(X)\ell(y)^{-1}.
\end{equation*}

\noindent and for all $s \in [0,1]$

\begin{equation*}
\nu(s) =q\left(F_{(1)}^{-1}(s)\right)\text{ }\mathrm{I}_(F^{-1}_{(1)}(s)\leq Z).
\end{equation*}

\noindent and\\

\begin{equation*}
f^{\ast\ast}_{s}(x)=f^{\ast}_{F^{-1}_{(1)}(s)}(x)=\mathrm{I}_(F^{-1}_{(1)}(s)\leq x), \ x \in \mathbb{R}.
\end{equation*}

\bigskip \noindent Finally, we suppose that the \textit{cdf} $F_{(1)}$ is increasing so that we have

\begin{equation}
F_{(1)}^{-1}(F_{(1)}(x))=x \text{ and } F_{(1)}(F_{(1)}^{-1}(s))=s, \text{ for } \ x\in \mathbb{R}, \  s\in (0,1). \label{HC}
\end{equation}

\bigskip

\noindent We have the following results for the asymptotic behavior of Takayama's statistic.

\begin{theorem} \label{theo1} \bigskip Let $0< \mathbb{E}(X^{2})<\infty$. Suppose that the regularity condition \ref{HC} holds. Then we have as $%
n\rightarrow \infty$
\begin{equation}
\sqrt{n}(T_{n}-T)=\mathbb{G}_{n,(1)}(g)+\beta _{n}(\nu) + o_{\mathbb{P}}(1), \label{TakayamaAsympRep}
\end{equation}

\bigskip \noindent with 
\begin{equation*}
\beta _{n}(\nu) = - \int_{0}^{1} \mathbb{G}_{n,(1)}(f^{\ast\ast}_{s}) \nu(s) ds.
\end{equation*}

\noindent We also have

\begin{equation}
\sqrt{n}(T_{n}-T)=\mathbb{G}_{(1)}(g)+\beta(\nu) + o_{\mathbb{P}}(1), \label{TakayamaAsympRep2}
\end{equation}

\bigskip \noindent with 
\begin{equation*}
\beta(\nu) = - \int_{0}^{1} \mathbb{G}_{(1)}(f^{\ast\ast}_{s}) \nu(s) ds.
\end{equation*}

\noindent In particular, we have
\begin{equation}
\sqrt{n}(T_{n}-T)\rightarrow \mathcal{N}(0,\sigma ^{2}), \label{TakayamaAsympDis}
\end{equation}

\noindent where 
\begin{equation}
\sigma ^{2}=\sigma _{1}^{2}+\sigma _{2}^{2}+2\sigma _{1,2} \label{TakayamaAsympVar}
\end{equation}

\noindent and 
\begin{equation*}
\sigma^2_1 = \int_{0}^{\infty} \left(g(x) - \mathbb{P}_X(g)\right)^2\,dF_{(1)}(x), \label{TakayamaAsympVar1}
\end{equation*}

\begin{equation}
\sigma^2_2 = \int_{[0,1]^2}\nu(s)\nu(t)(\min(s,t) - st)\,ds\,dt \label{TakayamaAsympVar2}
\end{equation}

\noindent and 
\begin{equation}
\sigma_{1,2} = \int_{0}^{1}\nu(s) \left( \int_{(x\leq F_{(1)}^{-1}(s))}
g(x)\,dF_{(1)}(x) - s\, \mathbb{P}_X(g)\right) \,ds. \label{TakayamaAsympVar3}
\end{equation}
\end{theorem}

\bigskip \noindent Before we begin the proof of this main theorem, the following lemma will allow us to make straightforward computations on formulas based on the functional empirical processes. (See \cite{wcia-srv-ang} , Chapter 5, more details on this lemma and other manipulations on $o_{\mathbb{P}}(c_n)$ with positive sequences $c_n$, $n\geq 1$).

\begin{lemma} \label{fep.lemma.tool.2} Let ($A_{n})$ and ($B_{n})$ be two sequences of real
valued random variables defined on the same probability space holding the
sequence $X_{1}$, $X_{2}$, etc.\\

\noindent Let $A$ and $B$ be two real numbers and Let $L(x)$
and $H(x)$ be two real-valued functions of $x\in \mathbb{R}$, with $(L,H)\in \mathcal{F}_{(1)}^2$.\\

\noindent Suppose that 
$$
A_{n}=A+n^{-1/2}\mathbb{G}_{n,(1)}(L)+o_{\mathbb{P}}(n^{-1/2})
$$ 

\noindent and 

$$
A_{n}=B+n^{-1/2}\mathbb{G}_{n,(1)}(H)+o_{\mathbb{P}}(n^{-1/2}).
$$ 

\noindent Then, we have

\begin{equation*}
A_{n}+B_{n}=A+B+n^{-1/2}\mathbb{G}_{n,(1)}(L+H)+o_{\mathbb{P}}(n^{-1/2}),
\end{equation*}

\noindent and

\begin{equation*}
A_{n}B_{n}=AB+n^{-1/2}\mathbb{G}_{n,(1)}(BL+AH)+o_{\mathbb{P}}(n^{-1/2})
\end{equation*}

\bigskip \noindent and if $B\neq 0$, we also have
\begin{equation*}
\frac{A_{n}}{B_{n}}=\frac{A}{B}+n^{-1/2}\mathbb{G}_{n,(1)}(\frac{1}{B}L-\frac{A}{%
B^{2}}H)+o_{\mathbb{P}}(n^{-1/2})
\end{equation*}
\end{lemma}

\bigskip

\noindent \textbf{Proof of Theorem \ref{theo1}}. Let us begin by recalling that 

\begin{equation}
T_{n}=1+\frac{1}{n}-\frac{2}{\mu _{n}n^{2}}\sum_{j=1}^{q}\left( n-j+1\right)
X_{j,n}.  \label{eq1}
\end{equation}

\noindent Let us denote
\begin{equation*}
A_{n}=\frac{1}{n^{2}}\sum_{j=1}^{q}\left( n-j+1\right) X_{j,n}.
\end{equation*}

\noindent Also, let $R_{n}=(R_{1,n},...,R_{n,n})$ be the rank statistic based on $X_{1},...,X_{n}$. We have
\begin{eqnarray*}
A_n &=& \frac{1}{n}\,\sum_{j=1}^{n}\,\left(1 - \frac{j}{n} \right)\,X_{j,n}\,%
\mathbf{1} \!\!\mathrm{I}_{(X_{j,n}\leq Z)} + \frac{1}{n^2}\sum_{j=1}^{n}\,X_{j,n}\,\,\mathbf{1} \!\!\mathrm{I}_{(X_{j,n}\leq Z)}\\
&=&\frac{1}{n}\,\sum_{j=1}^{n}\,\left( 1-\frac{R_{j,n}}{n}\right) \,X_{j,n\,}%
\mathbf{1}\!\!\mathrm{I}_{(X_{j,n}\leq Z)}+\frac{1}{n}\mathbb{P}_{n}\left(
\ell \right) ,
\end{eqnarray*}

\bigskip \noindent Let us define

\begin{equation}
\beta_{n}^{\ast}(q) = -\frac{1}{\sqrt{n}} \sum_{j=1}^{n} \left(\mathbb{G}_{n,r,(1)}(X_j)-F_{(1)}(X_j)
\right) q(X_j). \label{losallProc}
\end{equation}

\noindent \noindent This process has been introduced by Sall and Lo \cite{sall-lo}. It may be directly related to the functional empirical process  by using the Bahadur theorem (\cite{bahadur66}) as explained below. Let us use the representations given Subsection \ref{subsec12} of Section \ref{sec1}.\\

\noindent Now since $\mathbb{P}_{X}(\ell )$ is finite, we have $n^{-1}\mathbb{P}_{n}(\ell )=o_{\mathbb{P}}(n^{-1})$ and then,

\begin{eqnarray*}
A_{n}&=&\frac{1}{n}\sum_{j=1}^{n}\left( 1-\mathbb{G}_{n,r,(1)}(X_{j})\right) \ell(X_{j})+o_{\mathbb{P}}(n^{-1})\\
&=&\frac{1}{n}\sum_{j=1}^{n}\left( 1-F_{(1)}(X_{j})\right)\ell (X_{j}) +\frac{1}{n}\sum_{j=1}^{n}\left( F_{(1)}(X_{j})-\mathbb{G}_{n,r,(1)}(X_{j})\right) \ell (X_{j})+o_{\mathbb{P}}(n^{-1})\\
&=&\mathbb{P}_{n}\left( h\right) +\frac{1}{\sqrt{n}}\beta_{n}^{\ast}\left(
\ell\right) +o_{\mathbb{P}}(n^{-1}).
\end{eqnarray*}

\bigskip

\noindent Next, we obtain

\begin{equation}  \label{eqAn}
\sqrt{n}\left( A_{n}-\mathbb{P}_{X}(h)\right) =\mathbb{G}_{n,(1)}(h)+\beta_{n}^{\ast}\left( \ell \right) +o_{\mathbb{P}}(n^{-1/2}).
\end{equation}

\bigskip \noindent Finally the Takayama index can be written as

\begin{equation*}
T_{n}=1+\frac{1}{n}-\frac{2}{\mathbb{P}_{n}\left( I\right) }\left( \mathbb{P}%
_{n}\left( h\right) +\frac{1}{\sqrt{n}}\beta^{\ast}_{n}\left( \ell \right)
+o_{\mathbb{P}}(n^{-1})\right) .
\end{equation*}

\bigskip \noindent Let us go further. We recall that $T = 1 - \frac{2}{\mu} \mathbb{P}_X\left(h\right)$. We have

\begin{equation*}
\sqrt{n}\left( T_n - T \right) = -2 \left(\frac{\sqrt{n}\left(A_n - \mathbb{P%
}_{X}\left( h\right) \right) }{\mu_n} - \frac{\mathbb{P}_X\left(h\right)}{%
\mu\,\mu_n} \sqrt{n}\left(\mu_n - \mu \right)\right) + \frac{1}{n}.
\end{equation*}

\noindent But we also have $\sqrt{n}\left(\mu_n - \mu \right) = \mathbb{G}_{n,(1)}\left(I_d\right) $. From Equation (\ref{eqAn}), we get

\begin{equation*}
\sqrt{n}\left( T_n - T \right) = -\frac{2}{\mu_n}\left(\mathbb{G}_{n,(1)}(h)+\beta_{n}^{\ast}\left( \ell \right) +o_{\mathbb{P}}(n^{-1/2}) - \frac{\mathbb{P}%
_{X}\left( h\right)}{\mu} \mathbb{G}_{n,(1)}\left(I_d\right)\right) + o_p(1)
\end{equation*}

\bigskip \noindent By applying the last conclusion in Lemma \ref{fep.lemma.tool.2}, we arrive at
\begin{equation*}
\sqrt{n}\left( T_n - T \right)= -2 \left\{\mathbb{G}_{n,(1)}\left( \mu^{-1}\, h - \mathbb{P}_{X}\left(
h\right)\,\mu^{-2}\,I_d \right) + \mu^{-1}\,
\beta_{n}^{\ast}\left(\ell\right)\right\} + o_p(1).
\end{equation*}

\bigskip 
\noindent By using the definitions given above, in particular the definition of the function $q$, we may write

\begin{equation}
\sqrt{n}\left( T_n - T \right) = \mathbb{G}_{n,(1)}\left(g\right) +
\beta_{n}^{\ast}\left(q\right) + o_p(1). \label{takayamaEtap1}
\end{equation}

\bigskip \noindent By (\ref{quantilefunction}), we have

\begin{eqnarray*}
\beta^{\ast}_{n}(q) &=& - \frac{1}{\sqrt{n}} \sum_{j=1}^{n}  \left(\mathbb{G}_{n,r,(1)}(X_{j,n})-F_{(1)}(X_{j,n}) \right)\\
&=& - \frac{1}{\sqrt{n}} \sum_{j=1}^{n} n \int_{\frac{j-1}{n}}^{\frac{j}{n}} \left(\mathbb{G}_{n,r,(1)}(X_{j,n})-F_{(1)}(X_{j,n}) \right) q(X_{j,n})ds\\
&=&- \sqrt{n} \int_{0}^{1} \left(\mathbb{G}_{n,r,(1)}(F_{n,(1)}^{-1}(s))-F_{(1)}(F_{n,(1)}^{-1}(s)) \right) q(F_{n,(1)}^{-1}(s))ds.\\
\end{eqnarray*}

\noindent Now, by (\ref{QQFunction}), and next by (\ref{EEFunction}) and by Assumption \ref{HC}, we have

\begin{eqnarray*}
 \beta^{\ast}_{n}&=&- \sqrt{n} \int_{0}^{1} \left(\mathbb{G}_{n,r,(1)}(F_{(1)}^{-1}(\mathbb{V}_{n,(1)}(s)))-F_{(1)}(F_{(1)}^{-1}(\mathbb{V}_{(n),(1)}(s))) \right) q(F_{(1)}^{-1}(\mathbb{V}_{(n),(1)}(s))) ds \\
&=& -\sqrt{n} \int_{0}^{1} \left(\mathbb{U}_{n,(1)}(F_{(1)}(F_{(1)}^{-1}(\mathbb{V}_{n,(1)}(s))))-F_{(1)}(F_{(1)}^{-1}(\mathbb{V}_{n,(1)}(s))) \right) q(F_{n,(1)}^{-1}(s)) \\
&=& - \sqrt{n} \int_{0}^{1} \left( \mathbb{U}_{n,(1)}(\mathbb{V}_{n,(1)}(s))-\mathbb{V}_{n,(1)}(s) \right) q(F_{n,(1)}^{-1}(s)) ds\\
&=& - \int_{0}^{1} \sqrt{n} \left( s - \mathbb{V}_{n,(1)}(s)\right)q\left(F_{(1)}^{-1}\left(\mathbb{V}_{n,(1)}(s) \right)\right)\, ds\\
&-& \int_{0}^{1} \sqrt{n} \left( \mathbb{U}_{n,(1)}\left(\mathbb{V}_{n,(1)}(s)
\right) -s\right)\,q\left(F_{(1)}^{-1}\left(\mathbb{V}_{n,(1)}(s) \right)\right)\,
ds.
\end{eqnarray*}

\bigskip

\noindent From Shorack and Wellner \cite{shorackwellner}, page 511, we have for any $n\geq 1$,

\begin{equation*}
\sup_{0\leq s\leq 1} \left| \mathbb{U}_{n,(1)}\left(\mathbb{V}_{n,(1)}(s) \right)
-s \right| \leq \frac{1}{n}.
\end{equation*}

\noindent Thus, for $n\geq 1$,

\begin{equation*}
\beta_n(q) = - \int_{0}^{1} \sqrt{n} \left( s - \mathbb{V}_{n,(1)}(s)\right) q(F_{(1)}^{-1}(\mathbb{V}_{n,(1)}(s))) , ds+ o_{\mathbb{P}}(1)
\end{equation*}

\begin{equation*}
= - \int_{0}^{1}\gamma_{n,(1)}(s) q(F_{(1)}^{-1}(\mathbb{V}_{n,(1)}(s)))+o_{\mathbb{P}}(1).
\end{equation*}

\bigskip \noindent Here, we may use the Bahadur property  (See (\ref{bahadurRep}) in Subsection \ref{subsec12}, Section \ref{sec1}). We recall that 
by (\ref{HC}) and (\ref{empiricalprocessRealFonct}) in \ref{subsec12}, Section \ref{sec1}, we have $\alpha_{n,(1)}(s)=\mathbb{G}_{n,r,(1)}\left(F^{-1}_{(1)}(s)\right)$ and that $\nu(s) = q(F_{(1)}^{-1}(s))$. Next, we have

$$
\mathbb{G}_{n,r,(1)}\left(F^{-1}_{(1)}(s)\right)=\mathbb{G}_{n,(1)}(f^{\ast}_{F^{-1}_{(1)}(s)})=\mathbb{G}_{n,(1)}(f^{\ast\ast}_{s}),
$$

\noindent where, accordingly to the notation before Theorem \ref{theo1}, we simplified and wrote 

$$f^{\ast}_{F^{-1}_{(1)}(s)}=f^{\ast\ast}_{s}.
$$

\noindent We get

\begin{equation}
\beta_{n}(\nu) = \int_{0}^{1} \mathbb{G}_{n,(1)}(f^{\ast\ast}_{s})\,\nu(\mathbb{V}_{n,(1)}(s))\,ds +o_{\mathbb{P}}(1). \label{takayamaEtap2}
\end{equation}

\bigskip \noindent Now, we have 

\begin{eqnarray*}
\beta_{n}(\nu) &=& \int_{0}^{1} \mathbb{G}_{n,(1)}(f_s)\,\nu(s)\,ds\\
&+& \int_{0}^{1} \mathbb{G}_{n,(1)}(f^{\ast\ast}_{s})\,(\nu\left(\mathbb{V}_{n,(1)}(s)\right)- \nu(s))\,ds + o_{\mathbb{P}}(1)
\end{eqnarray*}

\noindent with

\begin{eqnarray*}
&&\left\vert \int_{0}^{1} \mathbb{G}_{n,(1)}(f^{\ast\ast}_{s})\,(\nu\left(\mathbb{V}_{n,(1)}(s)\right)- \nu(s))\,ds\right\vert \\
&\leq & \left( \sum_{s\in[0,1]} \left\vert \mathbb{G}_{n,(1)}(f_s)\right\vert\right) \int_{0}^{1} \left\vert\nu(\mathbb{V}_{n,(1)}(s)- \nu(s) \right\vert\,ds.
\end{eqnarray*}

\noindent Since $C_n=\sup_{s\in[0,1]} \left\vert \mathbb{G}_{n,(1)}(f_s)\right\vert$ weakly converges to $\sup_{s\in[0,1]} \left\vert \mathbb{G}_{(1)}(f_s)\right\vert$, which is an \textit{a.s.} finite random variable (the supremum of the Brownian bridge on $[0,1]$ is bounded in probability), we have that the sequence $C_n$ is bounded in probability (See lemma 8, Chapter 5, \cite{wcia-srv-ang}, page 120). Next  
$D_{n}(s)=\left\vert\nu\left(\mathbb{V}_{n,(1)}(s)\right)- \nu(s) \right\vert$ almost-surely  converges to zero form the \textit{a.s.} convergence of $\sup_{s\in[0,1]} \left\vert \mathbb{V}_{n,(1)}(s)-s\right\vert$ to zero. Since 

$$
D_{n}(s) \leq  2 \left|\frac{F_{(1)}^{-1}(Z)}{\mathbb{P}_X(I_d)}\right|,
$$
 
\noindent we may apply the Lebesgue Convergence Theorem to have, as $n\rightarrow +\infty$,

$$
\left( \sup_{s\in[0,1]} \left\vert \mathbb{G}_{n,(1)}(f_s)\right\vert \right) \int_{0}^{1} \left\vert\nu\left(\mathbb{V}_{n,(1)}(s)\right)- \nu(s) \right\vert\,ds \rightarrow_{\mathbb{P}} 0.
$$

\noindent Finally, by putting together the previous facts, we have
$$
\sqrt{n}(T_n-T)=\mathbb{G}_{n,(1)}(g)+\int_{0}^{1} \mathbb{G}_{n,(1)}(f^{\ast\ast}_{s})\,\nu(s)\,ds+o_{\mathbb{P}}(1).
$$

\noindent This is the representation (\ref{TakayamaAsympRep}).\\

\noindent A simple argument based on Riemann sums along with weak law criteria using characteristic functions yields

$$
\sqrt{n}(T_n-T)=\mathbb{G}_{(1)}(g)+\int_{0}^{2} \mathbb{G}_{(1)}(f^{\ast\ast}_{s})\,\nu(s)\,ds+o_{\mathbb{P}}(1).
$$
 
\noindent It is clear that $\sqrt{n}(T_n-T)$ is asymptotically Gaussian $\mathcal{N}(0,\sigma^1)$ since the couple $(\mathbb{G}_{(1)}(g), \int_{0}^{1} \mathbb{G}_{(1)}(f^{\ast\ast}_{s})\,\nu(s)\,ds)$ is normal. The computation of the variance-covariance of this vector is straightforward and is given below. We have \\

\begin{eqnarray*}
\sigma^2 &=& \mathbb{E}\left(\mathbb{G}_{(1)}(g)^2\right) + \mathbb{E}%
\left(\beta(\nu)^2 \right) + 2\, \mathbb{E}\left(\mathbb{G}_{(1)}(g)\,\beta(\nu)
\right)\\
&=:& \sigma^2_1 + \sigma^2_2 + 2\, \sigma_{1,2}.
\end{eqnarray*}

\noindent where $=:$ stands for the definition of the three terms of the left-hand member in the latter equality as 
$\sigma^2_1$, $\sigma^2_2$ and $2 \sigma_{1,2}$.  We easily find that

\begin{equation*}
\sigma^2_1 = \int_{0}^{\infty} \left(g(x) - \mathbb{P}_X(g)\right)^2\,dF_{(1)}(x).
\end{equation*}

\bigskip
\noindent Next, by a well-known formula, we have

\begin{equation}  \label{sigmanu}
\sigma^2_2=\mathbb{E}\left(\beta(\nu) \beta(\nu^{\prime}) \right) =
\int_{(0,1)^2}\nu(s)\nu(t)(\min(s,t) - st)\,ds\,dt.
\end{equation}

\noindent Concerning $\sigma_{1,2}$, we have

\begin{equation*}
\mathbb{E}\left(\mathbb{G}_{(1)}(g)\,\beta(\nu) \right) = \mathbb{E}\left(\mathbb{G}_{(1)}(g)\,\int_{[0,1]}\mathbb{G}_{(1)}(f^{\ast\ast}_{s}) \nu(s)\,ds \right).
\end{equation*}

\noindent Fubini's theorem implies that

\begin{equation*}
\mathbb{E}\left(\mathbb{G}_{(1)}(g)\,\beta(\nu) \right) = \int_{[0,1]}\, \nu(s) 
\mathbb{E}\left(\mathbb{G}_{(1)}(g)\mathbb{G}_{(1)}(f^{\ast\ast}_{s}) \right)\,ds.
\end{equation*}

\noindent But, we remark that

\begin{equation*}
\mathbb{E}\left(\mathbb{G}_{(1)}(g)\mathbb{G}_{(1)}(f^{\ast\ast}_{s}) \right)\,ds = \mathbb{P}%
_X\left(g\, f^{\ast\ast}_{s}\right) - \mathbb{P}_X(g) \, \mathbb{P}_X(f^{\ast\ast}_{s}),
\end{equation*}

\begin{equation*}
\mathbb{P}_X\left(g\, f^{\ast\ast}_{s}\right) = \int_{\mathbb{R}^*_+} \mathbf{1} \!\!%
\mathrm{I}_{(x \leq F_{(1)}^{-1}(s))} g(x)\, dF_{(1)}(x)
\end{equation*}

\noindent We finally get

\begin{equation*}
\sigma_{1,2} = \int_{0}^{1} \nu(s) \left( \int_{(x\leq F_{(1)}^{-1}(s))}
g(x)\,dF_{(1)}(x) - s\, \mathbb{P}_X(g)\right) \,ds.
\end{equation*}

\noindent This completes the proof of the Theorem \ref{theo1}. $\blacksquare$.

\bigskip

%%%%%%%%%%%%%%%%%%%%%%%%%%%%%%%%%%%%%%%%%%%%%%%%%%%%%%%%%%%%%%%
%%%%%%%%%%%%%%%%%%%%%%%%%%%%%%%%%%%%%%%%%%%%%%%%%%%%%%%%%%%%%%%

\section{Statistical of the default of decomposability} \label{sec3}

\subsection{Introduction}

In this section, we are concerned with the statistical estimation of the
decomposability gap of the Takayama statistic. This statistic is
surely non decomposable in the classical definition of Welfare analysts.
This study comes as a continuation of the works of Haidara and Lo who first
considered such an estimation. The reader is then referred to \cite{haidara-lo}
for a general introduction on this topic.\newline

\noindent It is remarkable that the results of Haidara and Lo extend to Takayama's statistic although they used indices based on the relative
poverty gaps. The reason is that they derived their estimation from the representation (\ref{TakayamaAsympRep}). This means that such results hold whenever that representation holds. \newline

\noindent In that sense, the coming theorem is a consequence of Formula (\ref{TakayamaAsympRep}) and Theorem 1 in \cite{haidara-lo}.\\

\noindent As a result, we will focus on the data driven applications and on comparison results with the Sen measure.
In this context, we will rephrase the statistical decomposability gap problem. We did this in the introduction with the deterministic index. We are going to describe it with the random index.\\

\noindent Now, suppose that the population is divided into $K$ subgroups $S_{1},...,S_{K}$ and for each $i\in \{1,...,K\}$, let us denote the subset
of the random sample $\{X_{1},...,X_{n}\}$ coming from $S_{i}$ by $\mathcal{E}_{i}=\{X_{i,1},...,X_{i,n_{i}}\}$ and then put $T_{n_{i}}(i)=T(X_{i,1},...,X_{i,n_{i}})$ the Takayama statistics on the $i^{th}$ subgroup. The decomposability gap is defined by
\begin{equation*}
gd_{n}=T_{n}-\frac{1}{n}\sum_{i=1}^{K}n_{i}T_{n_{i}}(i).
\end{equation*}

\noindent At this step, we have to precise our random drawing. We are going to use a probability
space in the form ($\Omega _{1}\times \Omega _{2},\mathcal{A}_{1}\otimes \mathcal{A}_{2},%
\mathbb{P}_{1}\otimes \mathbb{P}_{2})$ with $\mathbb{P=P}_{1}\otimes 
\mathbb{P}_{2}.$ We draw the observations in the following way. In each
trial, we draw a subgroup, the $ith$ subgroup $(\mathcal{E}_{i})$\ having
the occurring probability $p_{i}$. We define  

\begin{equation*}
\pi _{i,j}(\omega _{1})=1_{(\text{the }i^{th}\text{ subgroup is drawn at the 
}j^{th}\text{ trial})}(\omega _{1}),
\end{equation*}

\bigskip \noindent where, $1\leq i\leq K,1\leq j\leq n$. Now, given that the $i^{th}$
subgroup is drawn at the $j^{th}$ trial, we pick one individual in this
subgroup and observe its income $X_{j}(\omega _{1},\omega _{2}).$ We then
have the observations 
\begin{equation*}
\{X_{j}(\omega _{1},\omega _{2}),\text{ }1\leq j\leq n\}.
\end{equation*}

\noindent We have these simple facts. First, \ for $1\leq i\leq K,$ 
\begin{equation}
n_{i}^{\ast }=\sum_{j=1}^{n}\pi _{i,j}.
\end{equation}

\noindent Let us denote the distribution of $X_{j}$ given $(\pi _{i,j}=1)$, by  $F_{i,(1)}$ that is 
\begin{equation*}
\mathbb{P}(X_{j}\leq y\text{ }\diagup \pi _{i,j}=1)=F_{i,(1}(x).
\end{equation*}

\noindent We simply put $F_{i,(1}(x)=F_{i}$, $y\in \mathbb{R}$, to keep the notation simple. Then we have  
\begin{eqnarray*}
\forall (x\in \mathbb{R)},\mathbb{P}(X_{j}\leq y\text{ })&=&\sum_{i=1}^{K}\mathbb{P}(\pi _{i,j}=1)\mathbb{P}(X_{j}\leq y\text{ }\diagup \pi _{i,j}=1)\\
&=&\sum_{i=1}^{K}p_{i}F_{i}(x).\\
\end{eqnarray*}

\noindent We conclude that $\{X_{1},...,X_{n}\}$ is an independent sample drawn from $%
F_{(1)}(x)$ $=\sum_{i=1}^{K}p_{i}F_{i}(x)$, which is the mixture of the distribution
functions of the subgroups incomes.\\

\noindent Finally, we readily see that
conditionally on $n^{\ast }\equiv (n_{1}^{\ast },n_{2}^{\ast
},...,n_{K}^{\ast })=(n_{1},n_{2},...,n_{K})\equiv \overline{n}$ with $%
n_{1}+n_{2}+...+n_{K}=n,$ $\{X_{i,j},$ $1\leq j\leq n_{i}^{\ast }\}$ are
independent random variables with distribution function $F_{i}$.

\subsection{Notation}

\bigskip

Given all the previous preliminaries, we are able to state similar results of \cite{haidara-lo}. Denote for each subgroup $i$ $(1\leq i\leq K)$ 
\begin{equation*}
g_{i}(x)=2\mathbb{E}^{-2}(X^{i})I_{i}x-2\mathbb{E}^{-1}(X^{i})\left(
1-F_{i}(x)\right) x\mathbf{1}\!\!\mathrm{I}_{\left( x<Z\right) }
\end{equation*}

\noindent and
\begin{equation*}
\nu _{i}(x)=-2\mathbb{E}^{-1}(X^{i})x\mathbf{1}\!\!\mathrm{I}_{\left(
x<Z\right) }
\end{equation*}

\noindent Finally introduce as in \cite{haidara-lo}, 
\begin{equation*}
A_{1}=\sum_{i=1}^{K}p_{i}\left\{
\int_{0}^{1}(g-g_{i})^{2}(F_{i}^{-1}(t))dt-\left(
\int_{0}^{1}(g-g_{i})(F_{i}^{-1}(t))dt\right) ^{2}\right\} ,
\end{equation*}

\begin{equation*}
A_{2}=\sum_{i=1}^{K}p_{i}\int_{0}^{1}\int_{0}^{1}(s\wedge t-st)(p_{i}\nu
-\nu _{i})(F_{i}^{-1}(s))(p_{i}\nu -\nu _{i})(F_{i}^{-1}(t))dsdt,
\end{equation*}

\begin{equation*}
A_{31}=\sum_{i=1}^{K}p_{i}^{2}\sum_{h\neq
i}^{K}p_{h}\int_{0}^{1}\int_{0}^{1} \left[ {F_{h}(F_{i}^{-1}(s))\wedge
F_{h}(F_{i}^{-1}(t))}\right.
\end{equation*}

\begin{equation*}
\left. {-F_{h}(F_{i}^{-1}(s))F_{h}(F_{i}^{-1}(t))}\right] \newline
\nu (F_{i}^{-1}(s))\nu (F_{i}^{-1}(t))dsdt,
\end{equation*}

\begin{equation*}
A_{32}=\sum_{i=1}^{K}p_{i}\sum_{j\neq i}^{K}p_{j}\sum_{h\notin
\{i,j\}}^{K}p_{h}\int_{0}^{1}\int_{0}^{1}\left[ {F_{h}(F_{i}^{-1}(s))\wedge
F_{h}(F_{j}^{-1}(t))}\right.
\end{equation*}

\begin{equation*}
\left. {-F_{h}(F_{i}^{-1}(s))F_{h}(F_{j}^{-1}(t))}\right] \nu
(F_{i}^{-1}(s))\nu (F_{j}^{-1}(t))dsdt,
\end{equation*}

\begin{equation*}
B_{1}=\sum_{i=1}^{K}p_{i}\int_{0}^{1}\left\{ {%
\int_{0}^{s}(g-g_{i})(F_{i}^{-1}(t))dt}\right.
\end{equation*}

\begin{equation*}
\left. {-s\int_{0}^{1}(g-g_{i})(F_{i}^{-1}(t))dt}\right\} (p_{i}\nu -\nu
_{i})(F_{i}^{-1}(s))ds,
\end{equation*}

\begin{equation*}
B_{2}=\sum_{j=1}^{K}p_{j}\sum_{i\neq
j}^{K}p_{i}\int_{0}^{1}\int_{0}^{1}[s\wedge
F_{i}(F_{j}^{-1}(t))-sF_{i}(F_{j}^{-1}(t))],
\end{equation*}

\begin{equation*}
\times (p_{i}\nu -\nu _{i})(F_{i}^{-1}(s))\nu (F_{j}^{-1}(t))dsdt,
\end{equation*}

\begin{equation*}
B_{3}=\sum_{j=1}^{K}p_{j}\sum_{i\neq j}^{K}p_{i}\int_{0}^{1}\newline
\left\{ {\int_{0}^{F_{i}(F_{j}^{-1}(s))}(g-g_{i})(F_{i}^{-1}(t))dt}\right.
\end{equation*}

\begin{equation*}
\left. {-F_{i}(F_{j}^{-1}(s))\times \int_{0}^{1}(g-g_{i})(F_{i}^{-1}(t))dt}%
\right\} \nu (F_{j}^{-1}(s))ds,
\end{equation*}%
\begin{equation*}
gd=T(F_{(1)})-\sum_{i=1}^{K}p_{i}T(F_{i});\text{ }gd_{0,n}=T(F_{(1)})-%
\sum_{i=1}^{K}(n_{i}/n)T(F_{i})
\end{equation*}

\subsection{The theoretical result}

We have the following result.

\begin{theorem} \label{theo2} Let $\mathbb{E}X^{2}<\infty$ and for each $i \in {1,...,K}$,
$$
0 < \int x \ dF_{(1)}(x) \ dx <+\infty
$$

\noindent and, $F_{(1)}$ and each $F_{i}$, $1\leq i \leq K$ are increasing so that they are invertible.\\

\noindent Then we have
$$
gd_{n,0}^{\ast }=\sqrt{n}(gd_{n}-gd_{0})\leadsto \mathcal{N}(0,\vartheta _{1}^{2}+\vartheta_{3}^{2})
$$

\noindent and \\

$$
gd_{n}^{\ast }=\sqrt{n}(gd_{n}-gd)\leadsto \mathcal{N}(0,\vartheta _{1}^{2}+\vartheta _{2}^{2})
$$ 

\noindent with 
\begin{equation*}
\vartheta _{1}^{2}=A_{1}+A_{2}+A_{3}+2(B_{1}+B_{2}+B_{3})
\end{equation*}

\noindent and 

\begin{equation*}
\vartheta _{2}^{2}=\sum_{h=1}^{K}F_{h}{}^{2}p_{h}-\left(
\sum_{h=1}^{K}F_{h}p_{h}\right) ^{2}
\end{equation*}

\noindent for $F_{h}=\mathbb{E}g(X^{h})-J(F_{h})+\sum_{i=1}^{K}p_{i}\mathbb{E}%
F_{h}(X^{i})\nu (X^{i}),$ and

\begin{equation*}
\vartheta _{3}^{2}=\sum_{h=1}^{K}M_{h}{}^{2}p_{h}-\left(
\sum_{h=1}^{K}M_{h}p_{h}\right) ^{2}
\end{equation*}

\noindent for $M_{h}=\mathbb{E}g(X^{h})+\sum_{i=1}^{K}p_{i}\mathbb{E}F_{h}(X^{i})\nu
(X^{i}).$
\end{theorem}

\bigskip \noindent \textbf{Proof}. Based on Formula (\ref{TakayamaAsympRep}), the proof ofTheorem 1 in \cite{haidara-lo} applies line by line. $\blacksquare$.\\

\section{Datadriven applications}

\noindent \textbf{A - ESAM 1 Database, 1996}.\\

\noindent We consider the Senegalese database ESAM 1 of 1996 which includes 3278 households. We first consider the geographical decomposition into the areas (Dakar is the Capital). We have the Takayama measure values for the whole Senegal and for its ten sub-areas. The FAO scale has been used to obtain the equivalence-adult income for the households and poverty line has been taken equal to 143080 local monetary units (\textit{CFA franc CFA}). $ $\\

\begin{center}
\begin{tabular}{lcccccc}
\hline
Area & Senegal & Kolda & Dakar & Diourbel & St-Louis & Louga \\ \hline
Takayama (\%) & 93.14 & 78.57 & 96.51 & 86.65 & 93.92 & 88.59 \\ \hline
Size & 3278 & 198 & 1122 & 231 & 314 & 174 \\ \hline
\end{tabular}

$ $\\
\bigskip

\begin{tabular}{lccccc}
\hline
Area & Tambacounda & Kaolack & Thies & Fatick & Ziguinchor \\ \hline
Takayama (\%) & 80.81 & 89.10 & 88.24 & 87.37 & 94.60 \\ \hline
Size & 126 & 316 & 401 & 180 & 216 \\ \hline
\end{tabular}
\end{center}

\begin{figure}[tbph]
\includegraphics[width=.5\textwidth]{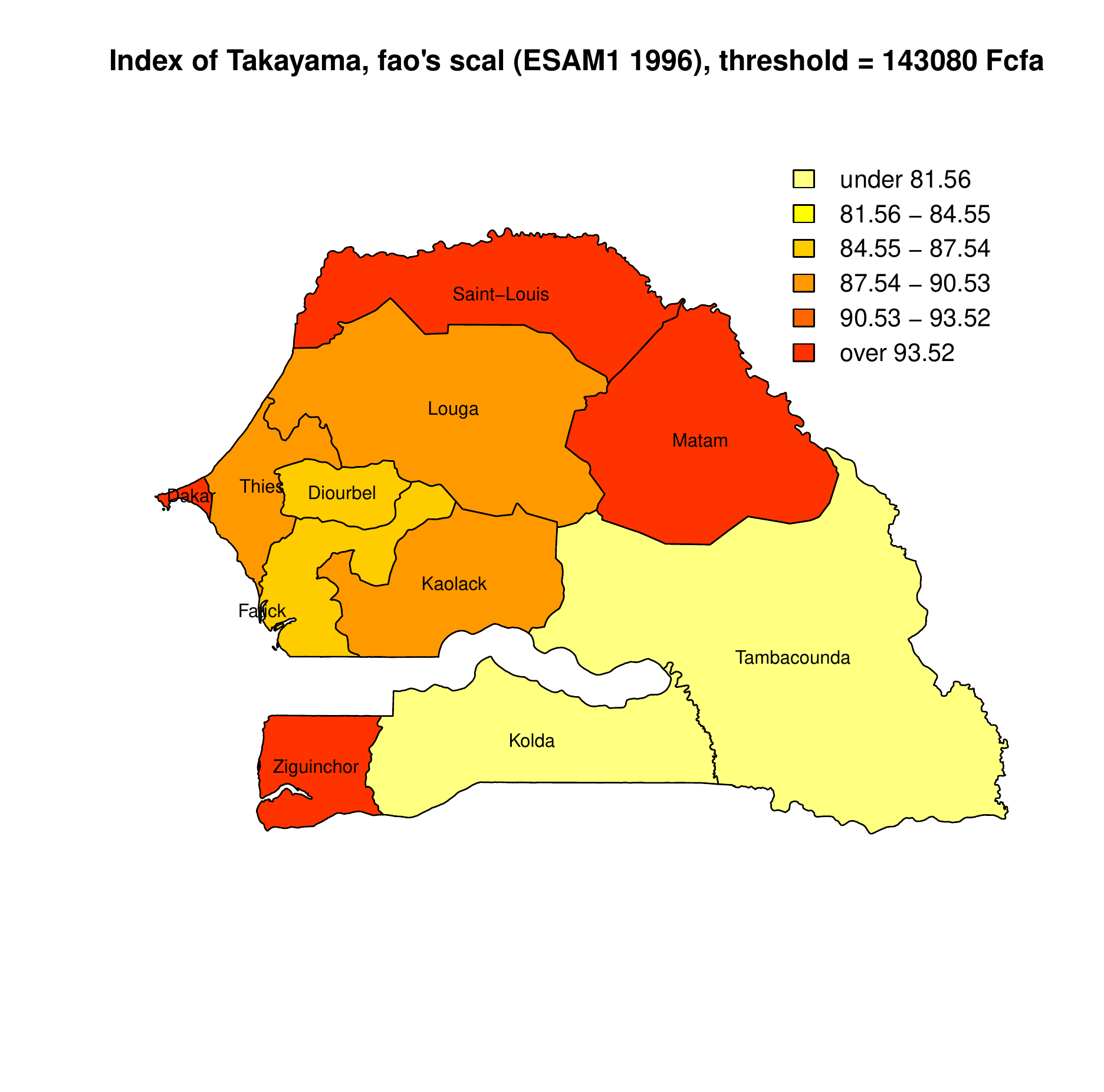}
\end{figure}

\noindent Let us compute the different variances $\vartheta _{1}^{2},\vartheta
_{2}^{2} $ and $\vartheta _{3}^{2}$ of Theorem \ref{theo2} with the
empirical estimations $p_{i}\approx n_{i}/n,$ . We obtain for the
geographical decomposability in Senegal : $\vartheta _{1}^{2}+\vartheta
_{2}^{2}=0.0834;$ $\vartheta _{1}^{2}+\vartheta _{3}^{2}$ $=0.0834$ and $%
gd_{n}=0.0203$. This gives the $95\%$ -confidence interval :

\begin{equation*}
dg\in \lbrack 0.0104;0.0302],\text{ } \sum_{i=1}^{k}\frac{n_{i}}{n}T_{n_{i}}(G_i)=0.9111,
\end{equation*}

\noindent that is 
\begin{equation}
T(F_{(1)})=0.9314 \in \lbrack 0.09215; 0.09413 \rbrack. \label{recompSenArea}
\end{equation}

\bigskip \noindent Now for a decomposition with respect to the household chief gender, we get the Takayama measure values.\\

\begin{center}
\begin{tabular}{lccc}
\hline
Gender & Senegal & Male & female \\ \hline
Takayama Index & 93.14\% & 93.66\% & 90.73\% \\ \hline
size & 3278 & 2559 & 719 \\ \hline
\end{tabular}
\end{center}

\bigskip
\noindent We get here $\vartheta _{1}^{2}+\vartheta _{2}^{2}= 2,4147.10^{-4}$ $%
\vartheta _{1}^{2}+\vartheta _{3}^{2}=2,4147.10^{-4}$, $gd_{n}=0.0124$
This gives the $95\%$ -confidence interval :

\begin{equation*}
dg\in \lbrack 0,9229, \ 0.9307], 
\end{equation*}

\noindent and
\begin{equation}
T(F_{(1)})=0.9314 \in \lbrack 0.93091; 0.93.198 \rbrack. \label{recompSenGender} 
\end{equation}

\noindent We get the conclusion that, in this case,  the gap of decomposability is not that low. Rather, it is statistically significant.\\

\bigskip \noindent \textbf{B - EPVC Database, 2004}.\\

\noindent We consider the Mauritanian database EPCV of 2004 which includes 9360
households. We first consider the geographical decomposition into the areas,
Nouakchott (Nktt) is the Capital. We have the Takayama measure values for
the whole Mauritania and for its thirteen sub-areas. The Oxforf scale has been used to obtain the equivalence-adult income for the households and poverty line has been taken equal to 94600 local monetary units (\textit{ougiya})\\

\begin{center}
\begin{tabular}{lcccc}
\hline
Area & Mauritania & Hod el Charghy & Hod el Gharby & Assaba \\ \hline
Takayama (\%) & 87.49 & 85.40 & 85.83 & 91.78 \\ \hline
Size & 9360 & 1211 & 469 & 514 \\ \hline
\end{tabular}

$ $\\
\bigskip

\begin{tabular}{lccccc}
\hline
Area & Gorgol & Brakna & Trarza & Adrar & Dakhlet Nouadhibou \\ \hline
Takayama (\%) & 75.76 & 77.93 & 84.01 & 88.82 & 98.79 \\ \hline
Size & 796 & 1190 & 1217 & 568 & 585 \\ \hline
\end{tabular}

$ $\\
\bigskip

\begin{tabular}{lccccc}
\hline
Area & Tagant & Guidimagha & Tiris Zemmour & Inchiri & Nktt \\ \hline
Takayama (\%) & 71.89 & 77.91 & 93.00 & 83.37 & 95.16 \\ \hline
Size & 490 & 234 & 284 & 205 & 1597 \\ \hline
\end{tabular}
\begin{figure}[tbph]
\includegraphics[width=.5\textwidth]{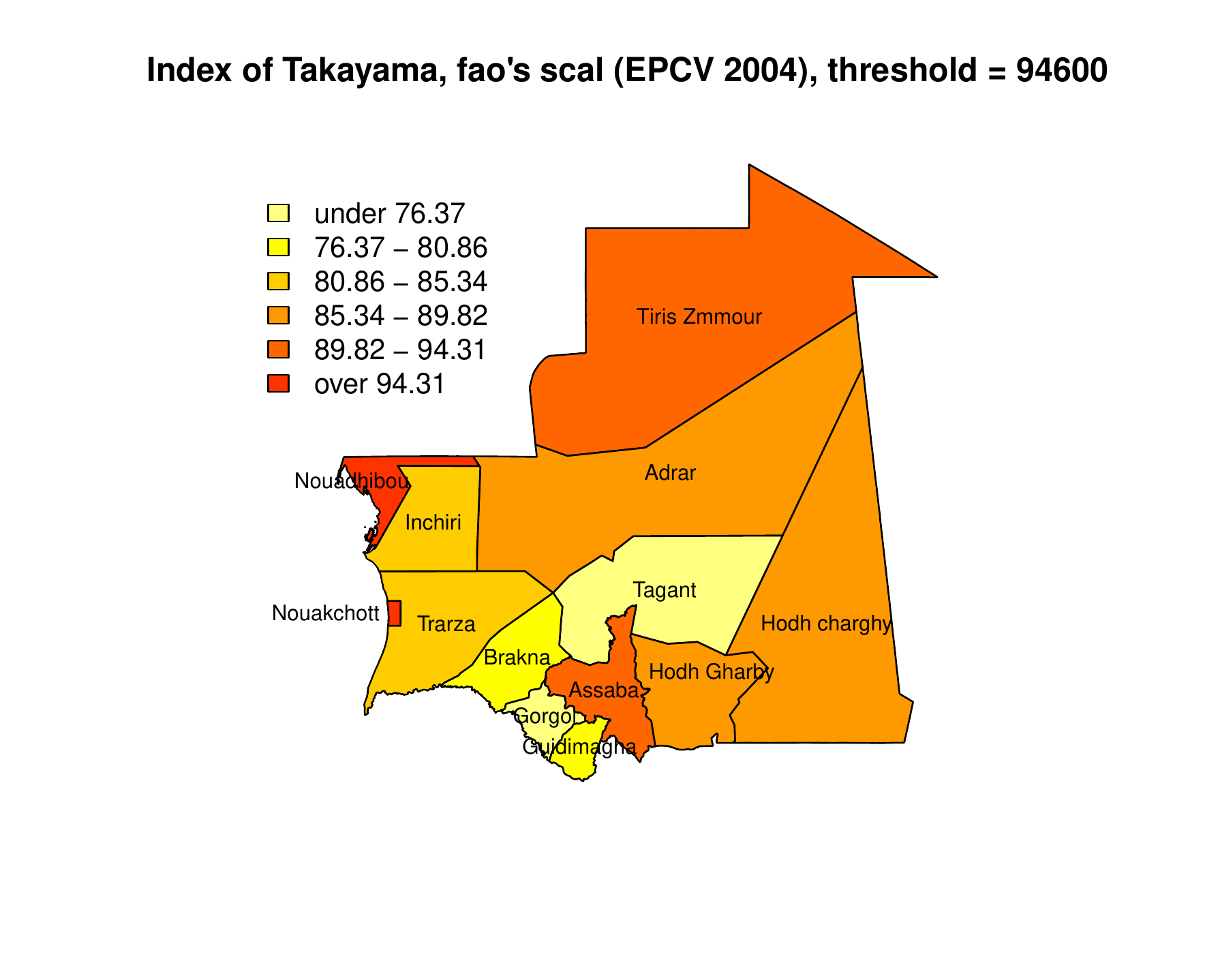}
\end{figure}
\end{center}

\noindent We obtain for the geographical decomposability in Mauritania: $\vartheta
_{1}^{2}+\vartheta _{2}^{2}=0.045295$ $\vartheta _{1}^{2}+\vartheta
_{3}^{2}=5,0970.10^{2}$ and $gd_{n}=0.0167$. This gives the $95\%$
-confidence interval :

\begin{equation*}
dg\in \lbrack 0.121; 0.0212],
\end{equation*}

\noindent and
\begin{equation}
T(F_{(1)})=0.8749\in \lbrack 0.8703, \ 0.8794 \rbrack. \label{recompMauriArea}
\end{equation}

\bigskip \noindent We get the conclusion that, in thees cases, the  decomposability gap is not significantly low.\\

\noindent \textbf{C - Analysis and comparisons}.\\

\noindent (a) In \cite{halo}, we have seen that decomposability gaps for the Sen index, have been estimated with confidence intervals with extreme lower and upper points not more far from zero that 1 to 9  per thousand, both for the gender and for the areas decompositions. We seize this opportunity to correct that paper and to say that there is no percentage in any confidence interval concerning the decomposability gap (dg). Instead, we have absolute numbers therein.\\

\noindent This conclusion is backed the empirical research in \cite{halo}, where the Sen index index has been observed as decomposable on the ESAM data.\\

\noindent (b) Compared to these indices, the Takayama index seems much less decomposable, from the statistical point of view. The gap of decomposability is statistically significant and are estimated at least at 0.7\%. But compared to the values of the Takayama's index, which turn around 80\%, the gaps are still relatively low.\\

\noindent And, in this case of significant lack of decomposition, our results may be used to recompose the global index in (\ref{recompSenArea}), (\ref{recompSenGender}) and (\ref{recompMauriArea}). And we see that decomposability gaps are low with respect to the values of the Takayama index values. We conclude that, based on the Senegal and Mauritania date, we may recommend to use the Takayama index as a decomposable one, at a statistical level.

\section{conclusion}
As in \cite{haidara-lo}, the Takayama's index which is theoretically non-decomposable has been observed as practically a decomposable one, based on the available data. But more importantly, the asymptotic law decomposability gap has been entirely described and the decomposability gap has been confined in 95\%-confidence intervals. This was possible because of the asymptotic representation of the Takayama's with respect to the functional empirical process and the Lo and Sall residual empirical process. The conclusions obtained in this paper are recommendable to other databases studies.

\bigskip \noindent \textbf{Acknowledgment}
The fourth (1 \& 4 \& 3) author acknowledges support from the World Bank Excellence Center (CEA-MITIC) that is continuously funding his research activities from starting 2014.\\

\end{document}